# The Enduring Effects of COVID-19 on Travel Behavior in the United States: A Panel Study on Observed and Expected Changes in Telecommuting, Mode Choice, Online Shopping and Air Travel

Mohammadjavad Javadinasr[1], Tassio B. Magassy[3], Ehsan Rahimi[1], Motahare (Yalda) Mohammadi[1], Amir Davatgari[1], Abolfazl (Kouros) Mohammadian[1], Deborah Salon[2], Matthew Wigginton Bhagat-Conway[4], Rishabh Singh Chauhan[1], Ram M. Pendyala[3], Sybil Derrible[1], Sara Khoeini[3]

1. Department of Civil, Materials, and Environmental Engineering, University of Illinois at Chicago, IL, USA;
2. School of Geographical Sciences and Urban Planning, Arizona State University, Tempe, AZ, USA;
3. School of Sustainable Engineering and the Built Environment, Arizona State University, Tempe, AZ, USA
4. Department of City and Regional Planning, University of North Carolina at Chapel Hill, Chapel Hill, North Carolina

**Abstract**

The explosive nature of Covid-19 transmission drastically altered the rhythm of daily life by forcing billions of people to stay at their homes. A critical challenge facing transportation planners is to identify the type and the extent of changes in people's activity-travel behavior in the post-pandemic world. In this study, we investigated the travel behavior evolution by analyzing a longitudinal two-wave panel survey data conducted in the United States from April 2020 to October 2020 (wave 1) and from November 2020 to May 2021(wave 2). Encompassing nearly 3,000 respondents across different states, we explored pandemic-induced changes and underlying reasons in four major categories of telecommute/telemedicine, commute mode choice, online shopping, and air travel. Upon concrete evidence, our findings substantiate significantly observed and expected changes in habits and preferences. According to results, nearly half of employees anticipate having the alternative to telecommute and among which 71% expect to work from home at least twice a week after the pandemic. In the post-pandemic period, auto and transit commuters are expected to be 9% and 31% less than pre-pandemic, respectively. A considerable rise in hybrid work and grocery/non-grocery online shopping is expected. Moreover, 41% of pre-covid business travelers expect to have fewer flights (after the pandemic) while only 8% anticipate more, compared to the pre-pandemic. Upon our analyses, we discuss a spectrum of policy implications in all mentioned areas.

# 1. Introduction

The Covid-19 pandemic has influenced nearly every facet of our lives, from daily social interaction to long-term residential location choices. The social and economic disturbances from lockdowns and travel restrictions caused many jobs to be shifted to a work-from-home setting to avoid social exposure and, thus, risk of getting or transmitting the novel coronavirus. As most Covid-19 restrictions are being or will be lifted in the U.S., many question the likelihood of which these changes will be permanent, and how the pandemic has impacted numerous areas in our society, especially working from home (WFH) and productivity (Galanti et al., 2021), telemedicine (Kichloo et al., 2020), commute mode choice (Shakibaei et al., 2021), online shopping (Kim, 2020), and air travel (Shamshiripour et al., 2020).

To assess the extent to which these societal changes will be long term phenomena, the COVID Future Research Team launched a multi-wave survey to collect U.S. residents' opinions about the pandemic, in addition to socioeconomic characteristics, lifestyle preferences, and mobility patterns during these unprecedented times (Chauhan et al., 2021). The panel survey is currently (August 2021) ongoing in multiple waves and for this study, two waves of the collected data, from April 2020 to October 2020 (wave 1) and from November 2020 to May 2021(wave 2) have been utilized. The survey approached pandemic-related behaviors before the pandemic, during the pandemic (capturing information about stay-at-home orders and other social restrictions), and post-pandemic expected behaviors (after all restrictions will be lifted, and the virus is no longer a threat).

Most published studies until now have used data from the first few months into the pandemic. Although important, at this period people were experiencing massive changes without having a clear picture of the future. This period is corresponding to the wave 1 data in our analyses. At the end of 2020 and the beginning of 2021, people had a more realistic evaluation of both the pandemic effects and the post-Covid future, especially with the hopes brought by the invention of Covid-19 vaccines. This period coincides with wave 2 data in our analyses. Moreover, despite the dynamic and ever-changing nature of occurrences and preferences during the pandemic, most studies have used cross-sectional data to investigate alterations in people's behaviors and habits. Our panel study provides the opportunity to scrutinize changes and their evolution through time and build insights upon more reliable findings by differentiating between transient and long-lasting changes. To extent of our knowledge, there has not been a comprehensive panel survey study that addresses different aspects of travel behavior changes caused by the Covid-19 pandemic in the U.S.

This research is set to explore the dynamics of people's activity-travel behavior before, during, and after the Covid-19 pandemic by focusing on four major categories of telecommuting (work from home), travel mode choice, online shopping, and air travel. We build our findings based on a unique nationally representative two-wave panel data, encompassing a sample size of 2,973 respondents, collected in the United States (U.S.). Using descriptive and inferential statistic measures, we investigate the *observed* changes from the pre-pandemic to wave 1, *observed* transitions from wave 1 to wave 2, and self-reported *expected* changes after the pandemic. Questions regarding the "pre-pandemic" period are asked in wave 1 and questions about expectations for the "post-pandemic" period are asked in wave 2 of the data collection. Moreover, our data and analyses include a range of underlying factors corresponding to observed/expected changes in all 4 categories. Exploring evidence from the revealed and the expected behaviors can be especially advantageous to shed light on the levels to which preferences and habits of U.S. adults have changed as well as projecting the stickiness of the changes after the pandemic ends. We note, however, that respondents here have actually lived the scenario given to them, which we believe, makes their responses more likely to apply than traditional stated-preference surveys. The remainder of this article is organized as follows. In Section 2, we present the literature review. In Section 3, we describe the data. In Section 4, we analyze and discuss the findings. Finally, in Section 5, we present the conclusions and policy implications.

## 2. Literature Review

The rapid human-to-human transmission of Covid-19 meant almost any human interaction risked exposure to infection. Accordingly, from the early days of the pandemic, leaving homes and going to offices became a risky behavior, which yielded a substantial increase in the number of telecommuters. Numerous studies have attempted to explore the effects of this mandatory and vast experience of WFH on employees' perception as well as transportation systems from different angles. Analyzing over 100,000 tweets expressing people's attitudes toward WFH experience during the pandemic in India, Dubey and Tripathi (2020) observed that more than 73% of people had a positive sentiment, and over 60% depicted trust and joy for WFH culture. According to a study conducted in Lithuania, sociodemographic variables such as gender, age, education, and work experience influence the efficiency and quality of WFH from the teleworkers' perspective (Raišienė et al., 2020). They reported the most satisfied teleworker group as the Millennial females with a graduate degree who have 4-10 years of experience in the management and administration field and WFH twice a week. Concerning the transportation network, by analyzing short-term reduction in money and time costs based on car and transit use, Hensher et al.,(2021b) stated that regardless of the stickiness of WFH in the future, sustainable telecommuting is an appealing policy that can decrease both congestions on the roads and crowding in public transport.

Using early data from the COVID Future Survey, Salon et al., (2021) explored the potential stickiness of pandemic-induced behavior changes by focusing on peoples' stated responses[1] regarding the post-covid future. According to the results, 26% of workers expected to telework at least a few times a week (after the pandemic), which is double the fraction of the same group before the pandemic. This increased number of teleworkers can decrease the car commute kilometers by approximately 15%. Moreover, they reported that transit commute trips are expected to decline by 40% after the pandemic is over.

One of the main factors affecting the predominance of telecommuting in the future is productivity. Some elements potentially can contribute to reducing the productivity of WFH such as distracting work environment, family-work conflict, and social isolation; whereas other elements can contribute to increasing productivity such as self-leadership and autonomy (Galanti et al., 2021).

The Covid-19 pandemic also engendered the need for health care facilities to rapidly develop and deploy telemedicine as a viable solution to address the overwhelming number of patients (Ortega et al., 2020). Telemedicine has the potential to provide health care through telecommunication technologies and can be an effective approach in terms of exchanging valid information for the diagnosis, reducing overcrowding in hospitals, and undertaking research and evaluation (Mahajan et al., 2020).

The pandemic also caused significant disruption in the transportation system. Many remote workers were not going to work anymore, or at least not as frequently as before the pandemic, and many adapted their commute mode choice to avoid virus exposure. For instance, during the lockdown in Greece, the trip frequency was decreased by almost 50%, although the trip duration was increased (Politis et al., 2021). A major decline in the utilization of all major public transportation modes in Turkey was reported (Shakibaei et al., 2021). Similar disruptions were also observed in the U.S., and the average commuting volume in terms of the total number of commuters within 24 hours in a given county across the country decreased almost by 65% compared to typical daily values (Klein et al., 2020). In the U.K., Harrington and Hadjiconstantinou, (2020) found that before the pandemic, 72% and 27% of the employed sample were car and transit commuters, respectively; of the car commuters, 81.9% plan to keep using their vehicles once restrictions are lifted, while 3.6% and 6.5% plan to change to walking and cycling, respectively; among transit users, 49.0% plan to switch mode. Overall, using shared modes, such as transit and ridesharing, has been undergone the most significant changes among other modes (Abdoli and Hosseinzadeh, 2021). This is mostly due to the high correlation with exposure risks which has resulted in a drop in shared mobility ridership (Rahimi et al., 2021).

---

[1] Please note that the these post-covid expectations were asked in early stages of the pandemic (i.e., wave 1), and are different than the post-covid expectations that are asked in wave 2 of the survey, which are presented in this study.

Significant changes were also found in shopping patterns. With the fear of being exposed to the virus from in-store purchases, many increased online shopping frequency instead (Ellison et al., 2021). Capturing changes in behaviors and perceptions of grocery shopping, Gerritsen et al., (2020) noted that 40% of household shoppers reduced the frequency of in-store grocery shopping. Recent literature has been discussing the extent to which shopping changes will remain in a post-pandemic life. Salon et al., (2021), reported around half of the new online shoppers stated they expect to continue to grocery shop online at least a few times a month. A study in Chicago revealed that around 74% of people stated they would rely more on online grocery shopping in the first few months after the pandemic, and 59% expressed willingness to order their groceries online even far after the pandemic (Shamshiripour et al., 2020). This study also reported that out of those who shopped online groceries at least once during the pandemic, 39% said it was their primary way of meeting their grocery needs during the height of the pandemic.

Another travel-related segment that was severely impacted by the pandemic restrictions was long-distance air travel. In 2020, global air passenger traffic experienced an overall reduction of about 60% compared to 2019 (ICAO, 2021). Moreover, airports will suffer the loss of more than 94 billion USD of revenue by the end of 2021, a drop in half of the revenue expectations compared to 2019 (ACI, 2021). Using early data from the COVID Future Survey to explore long-distance travel and the potential recovery of the airline industry, Conway et al., (2020) verified that frequent air travelers were more likely to expect a decrease in their air travel compared to the infrequent group. Most personal travelers expect changes due to the perceived risk of sharing close space with others, a concern that should fade out over time as the pandemic gets under control. On the other hand, most business travelers reported an expectation of a reduction in their air travel caused by an increased reliance on virtual communication. Discussing aviation-related impacts, Serrano and Kazda, (2020) believed that successful airports in the future may present a more tech-oriented setting, with biometric and self-service processes, reducing human-to-human interactions, as well as potential diversification in airport activities aiming at non-passenger revenue to compensate for declines in air travel.

## 3. Survey and Data Collection

We aimed to investigate changes in pandemic-related travel behaviors and preferences across time for the same population. Therefore, this paper uses the number of respondents present in both Wave 1 (responses collected from April 2020 to October 2020) and Wave 2 (responses collected from November 2020 to May 2021) of the COVID Future Panel Survey (Chauhan et al., 2021) including 2,973 observations in the United States.

The COVID Future Panel Survey is a longitudinal survey collecting travel-related behaviors and attitudes, covering various themes, such as lifestyle attitudes, travel patterns, telecommuting, telemedicine, online learning, and shopping, in addition to socioeconomic and demographic information. In addition to the questions regarding the "during pandemic" behaviors and attitudes, there are questions regarding the "pre-pandemic" (asked in wave 1) as well as questions about the "post-pandemic" (asked in wave 2). The survey is ongoing over multiple waves, and more details regarding its status and data availability can be found at https://covidfuture.org/.

To control for demographic discrepancies and provide more representative results, the panel data were weighted using PopGen 2.0 and were controlled for age, education level, gender, Hispanic status, household income, number of household vehicles, employment status, and white race based on American Community Survey (ACS) 5-year Estimates (U.S. Census Bureau, 2020). The sample was also divided into nine modified census divisions for weighting, and each subsample represents the marginal distribution of the region (Fig. 1). For more details about the weighting process, please refer to (Chauhan et al., 2021).

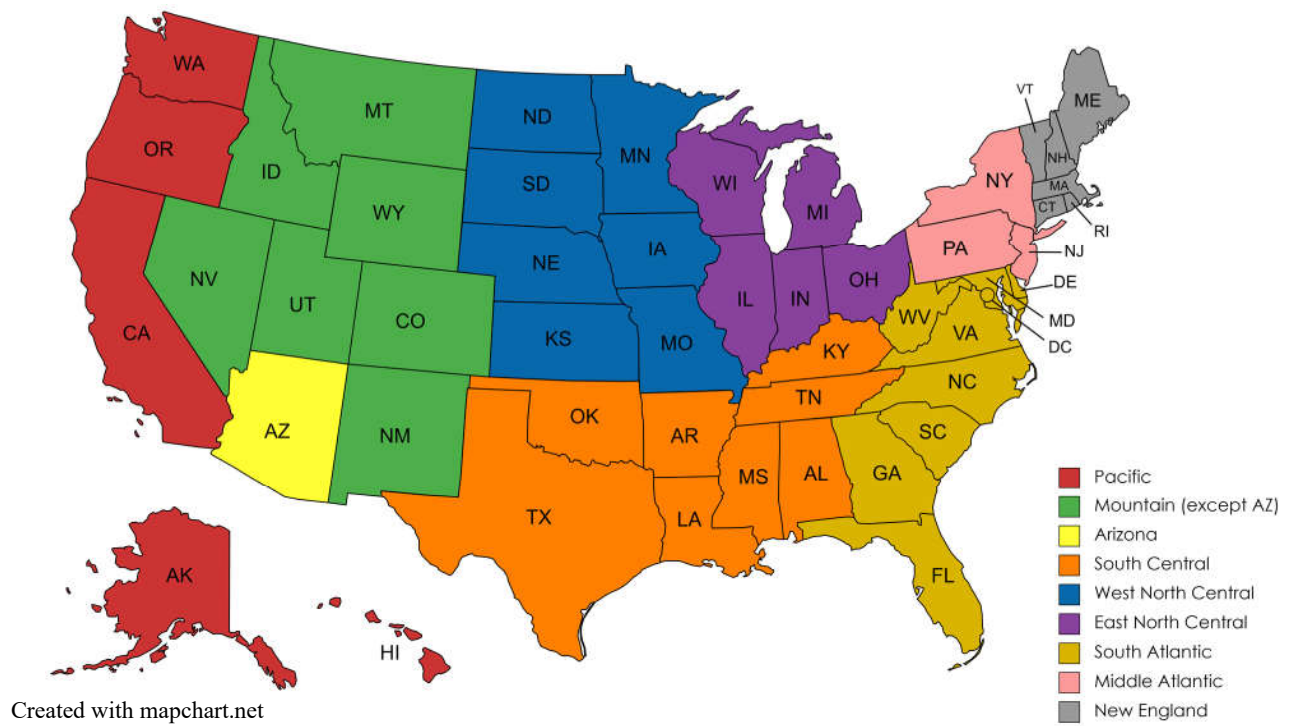

Created with mapchart.net

**Figure 1. Used census divisions for weighting the panel sample.**

After the weighing process, the distributions of the variables used for the weighting scheme match the national distributions (table 1). Nonetheless, not all possible variables were controlled for the weighted sample, thus some minor differences are observed when comparing some demographic variables and national proportions.

Table 1. Socioeconomic distribution of weighted and unweighted panel samples.

| | Unweighted Sample (N=2,973) | Weighted Sample (N=2,973) | Adults in the United States (2019) |
|---|---|---|---|
| *Age[a]* | | | |
| 18-29 | 8.7% | 21.0% | 21.0% |
| 30-44 | 22.2% | 25.2% | 25.2% |
| 45-59 | 25.8% | 24.4% | 24.4% |
| 60 years and above | 43.3% | 29.4% | 29.4% |
| *Gender[a]* | | | |
| Female | 64.3% | 51.3% | 51.3% |
| Male | 35.7% | 48.7% | 48.7% |
| *Education[a]* | | | |
| High school or less | 12.0% | 39.0% | 39.0% |
| Some college | 28.6% | 30.4% | 30.4% |
| Bachelor or higher | 59.4% | 30.6% | 30.6% |
| *Employment[a]* | | | |
| Employed | 57.1% | 62.0% | 62.0% |
| Non-employed | 42.9% | 38.0% | 38.0% |
| *Hispanic status[a]* | | | |
| Hispanic | 7.6% | 16.4% | 16.4% |
| Non-Hispanic | 92.4% | 83.6% | 83.6% |
| *Race[a]* | | | |
| Non-white | 14.2% | 26.4% | 26.4% |
| White | 85.8% | 73.6% | 73.6% |
| *Household size* | | | |
| 1 | 22.7% | 12.1% | 16.7% |
| 2 | 41.8% | 37.6% | 32.9% |
| 3 | 15.8% | 20.6% | 18.7% |
| 4+ | 19.6% | 29.7% | 31.7% |
| *Presence of children* | | | |
| Not present | 78.0% | 75.5% | 67.1% |
| Present | 22.0% | 24.5% | 32.9% |
| *Tenure* | | | |
| Homeowner | 69.2% | 63.8% | 65.7% |
| Not homeowner | 30.8% | 36.2% | 34.3% |
| *Household vehicles[a]* | | | |
| 0 | 7.2% | 9.3% | 9.3% |
| 1 | 37.9% | 22.6% | 22.6% |
| 2 | 40.4% | 37.4% | 37.4% |
| 3+ | 14.4% | 30.7% | 30.7% |
| *Household Income[a]* | | | |
| Low Income – Less than $49,999 | 21.8% | 18.9% | 18.9% |
| Medium Income – $50,000 to $99,999 | 44.1% | 41.1% | 41.1% |
| High Income – More than $100,000 | 34.1% | 40.0% | 40.0% |

Note. [a]Variables used on the weighting scheme

## 4. Results and Discussion

In this section, we present and discuss the important findings of the COVID Future Panel Survey, regarding the following categories: Telecommute, Productivity and Telemedicine, Commute Trips and Mode Choice, Online shopping, and Air Travel.

## 4.1. Telecommute, Productivity and Telemedicine

### 4.1.1. Telecommute

One of the strategic protective steps to break the chain of the coronavirus outbreak was to ask employees to work from home. This measure along with personal health concerns related to the risk of attending workplaces has made a remarkable change in the WFH patterns. The various aspects discussed in this section include the availability (i.e., having the option to WFH) among different population groups, the evolution of the adoption of working from home from the pre-pandemic to the (expected) post-pandemic, changes in productivity, and underlying factors.

Fig. 2 presents the availability of WFH across various population groups of workers who have not been laid off during the pandemic. As can be seen, the average percentage of the employees with the option to WFH has increased from 37% in pre-pandemic to 58% and 53% in waves 1 and 2, respectively. The transition from pre-pandemic to wave 1 indicates a remarkable surge across all population groups, yet it is not equally distributed. In all periods depicted in fig. 2, females, households with an annual income lower than $120 k and people with education levels

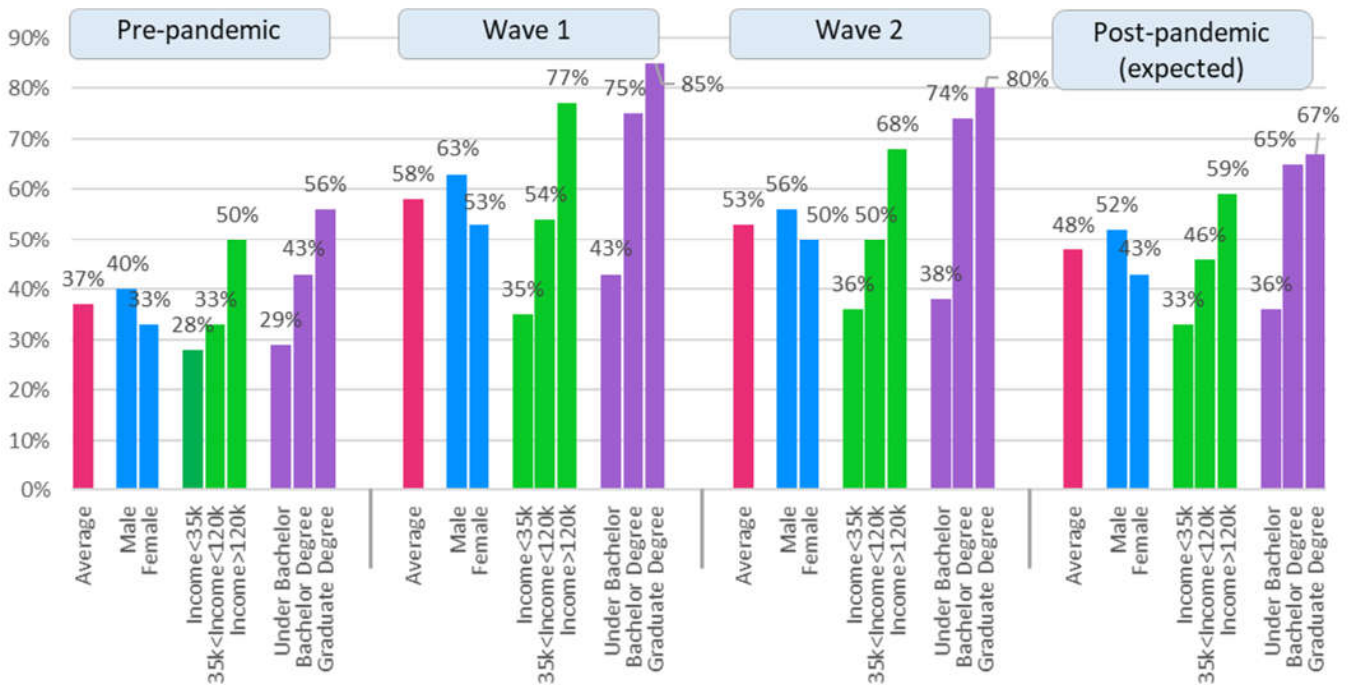

**Figure 2. The work from home availability across different demographic groups in pre-pandemic, wave 1, wave 2, and post-pandemic periods.**

lower than bachelor's degrees have below-average access to the opportunity to WFH. The two major contributing factors explaining the underlying heterogeneity are income and education level which are positively correlated with the WFH availability. There is a gender gap indicating that males have more options to WFH compared to females in all periods, yet the effect is not as considerable as income or education. After the pandemic, on average, 48% of the respondents expect to have the option to WFH, indicating a 30% growth compared to the pre-pandemic period.

Considering that having the option to WFH does not necessarily imply choosing it, we explored the dynamics of working from home by accounting for different levels of adoption. Based on the collected data, different workers are categorized into three classes of "frequent" (i.e., WFH more than once a week) or "infrequent" (i.e., WFH once a week or less) or "no option". Fig 3 illustrates the proportions and transitions of respondents who were employed before the pandemic and fall into one of these three classes. The first three columns correspond to the *observed* WFH behavior in pre-pandemic, wave 1, and wave 2 periods, respectively, whereas the last column refers to the *expected* WFH behavior after the pandemic. Moreover, table 2 presents the result of the statistical Z-tests on percentage changes in different classes of WFH. The results show that the fraction of employees without an option to WFH has decreased from 63% (pre-pandemic) to 36% and 42% in wave 1 and wave 2, respectively. The proportion of frequent telecommuters has substantially expanded from 16% (pre-pandemic) to 46% (i.e., 187% rise) in wave 1 which is also sustained through wave 2. This remarkable change has made an unprecedented opportunity to study the effects of working from home, on a vast scale, on different aspects of people's daily lifestyle, travel behavior, and the consequences on the transportation sector. Another observation is that through the transition from pre-pandemic to wave 1, 13% of the formerly employed respondents have been laid off. The unemployment rate then started to recover by reaching 9% in wave 2.

Regarding the post-covid situation, 52% of the respondents expect not to have the opportunity to WFH, which indicates a 17% reduction in this class compared to the pre-pandemic period. However, the most considerable change can be seen in the frequent commuters' class by increasing from 16% in pre-pandemic to 34% in post-pandemic (i.e., 112% growth).

One interesting finding is the notable drop in infrequent telecommuters by turning into the frequent category during and after the pandemic periods. The proportion of infrequent telecommuters is expected to be 14% after the pandemic manifesting a 33% drop compared to the pre-pandemic situation. This also implies that after having the mandatory experiment of WFH, previously infrequent telecommuters have become habituated to this *new normal* and expect to continue incorporating it more frequently in their daily lifestyles if they are given the option.

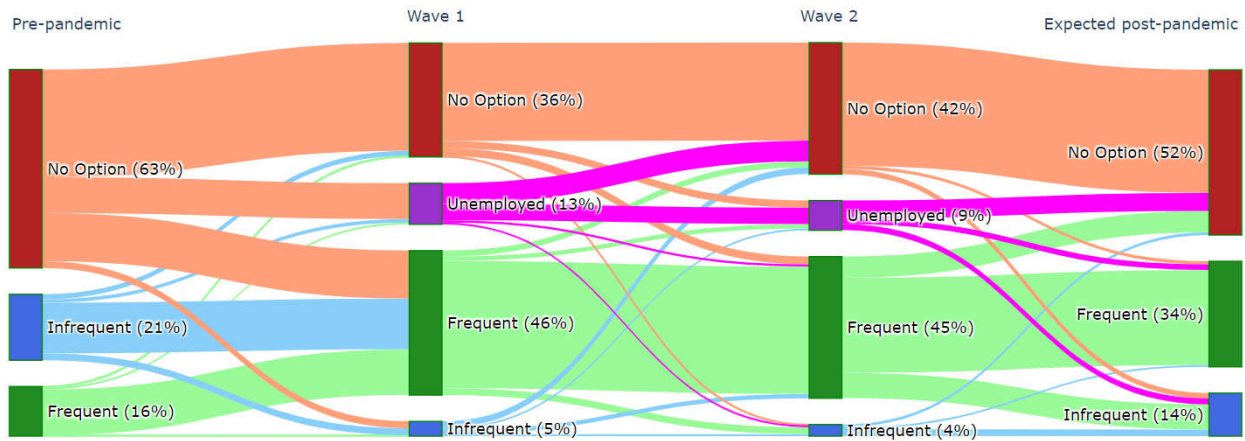

**Figure 3. Proportions and transitions of different levels of work from home adoption from pre-pandemic to post-pandemic. Please note that "Frequent" refers to WFH more than once/week, and "Infrequent" refers to WFH once/week or less.**

**Table 2. Percentage change and significance in different categories of WFH adoption from pre-pandemic to post-pandemic.**

| WFH level | Change in percentage (p-value) | | | |
|---|---|---|---|---|
| | Pre-pandemic → Wave 1 | Wave 1 → Wave 2 | Wave 2 → Post-pandemic | Pre-pandemic → Post-pandemic |
| No option | −43% *** (0.000) | +16% *** (0.000) | +24% *** (0.000) | −17 *** (0.000) |
| Frequent | +187% *** (0.000) | −2% (0.077) | −24% *** (0.000) | +112% *** (0.000) |
| Infrequent | −76% *** (0.000) | −20% ** (0.044) | +250% *** (0.000) | −33% *** (0.000) |

*** Indicates a significance level of 99%.
** Indicates a significance level of 95%.

### 4.1.2. Productivity

Considering that workers' intention to WFH after the pandemic can be highly correlated with their perceived productivity, we asked telecommuters in waves 1 and 2 to assess their productivity compared to the pre-pandemic period. Fig. 4 summarizes the result suggesting that around 30% of the respondents in both waves expressed higher productivity whereas the percentages of the individuals with lower perceived productivity were 24% in Wave 1, and then declined to 20% in wave 2. In general, the proportion of the respondents who believed their productivity has been higher or the same as the pre-pandemic period was 60% in wave 1 which increased to 71% in wave 2. Other than "higher", "same" and "lower" productivity options, respondents had the alternative to select "In some ways, it has increased and in other ways, it has decreased" which is denoted by the "Both – and +" in fig .4.

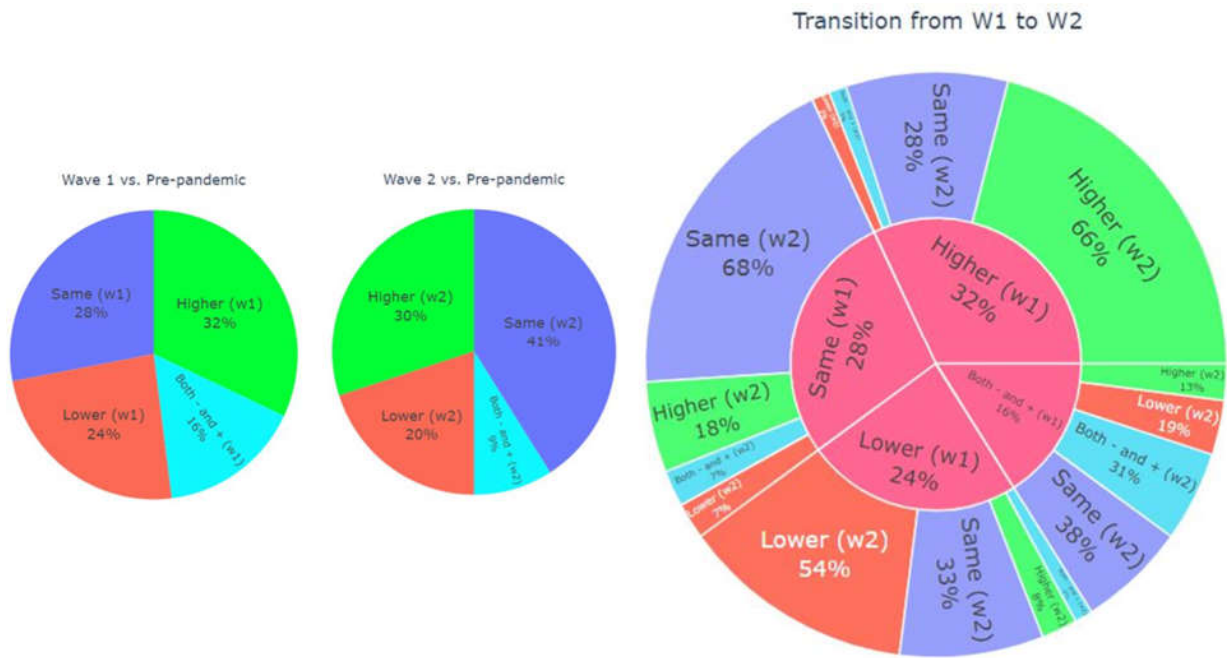

**Figure 4. The shares and transitions of different types of perceived productivity of telecommuters in wave 1 and wave 2, compared to the pre-pandemic period. Note that W1 refers to wave 1 and W2 refers to W2.**

To shed light on the underlying factors determining the perceived productivity of employees during the WFH period, we asked respondents to select among a set of provided reasons with the ability to choose more than just one item. Fig. 5(a), and 5(b) show the proportions of the selected components corresponding to lower and higher productivities, respectively. Accordingly, "more distractions at home" is the number one negative factor in both waves causing lower productivity (selected by 59% and 57% in waves 1 and 2, respectively) followed by "feeling sad, depressed, or burned out" that was only asked in wave 2 (selected by 42%). One notable observation here is that the frequency of the two factors of "need equipment or technology not available at home" and "lack of comfortable workspace" has decreased by 9% from wave 1 to wave 2, suggesting that people and companies have learned to adapt to the new work situation by creating a more conducive environment to WFH.

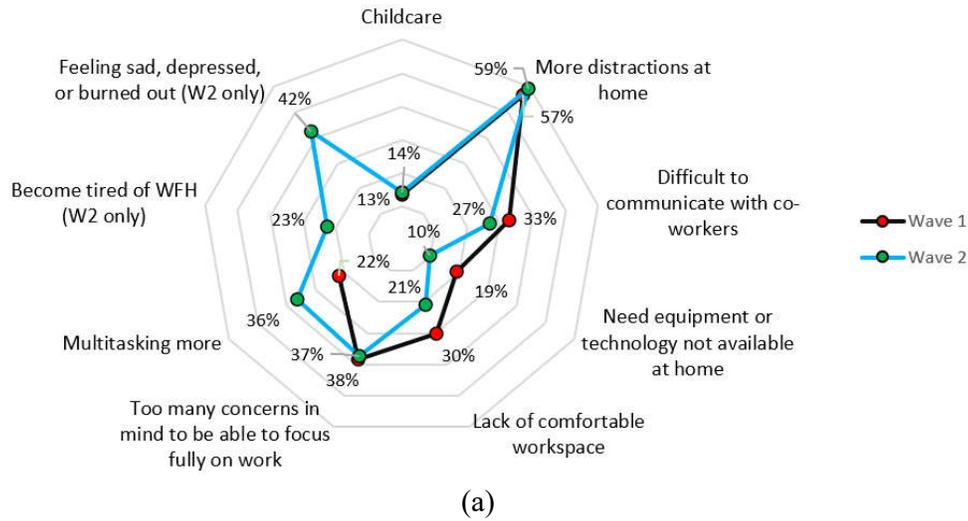

(a)

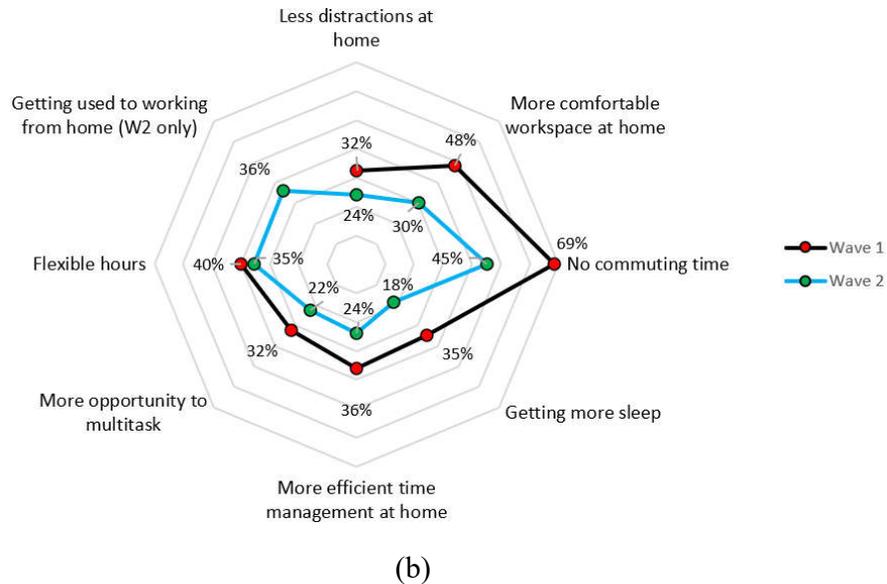

(b)

**Figure 5. Different factors affecting WFH productivity for respondents who reported (a) lower and (b) higher perceived productivity (compared to the pre-pandemic situation) reported in wave 1 and wave 2.**

On the other hand, the first selected positive factor associated with being more productive in both waves is "no commute" (selected by 69% and 45% in waves 1 and 2, respectively) followed by "more comfortable workspace at home" (selected by 48%) in wave 1 and "flexible hours" (selected by 35%) in wave 2.

### 4.1.3. *Telemedicine*

Telemedicine platforms can benefit both health care facilities by alleviating overcrowding and also low-acuity patients by providing medical guidelines without disease exposure risks (Rockwell et al., 2020). Our data shows that since the beginning of the pandemic until wave 2 of the survey, 2228 individuals have attended medical appointment(s) among which 51% have incorporated either an online or a mix of online and in-person arrangement(s) (fig. 6). This ratio demonstrates the unprecedented high level of telemedicine adoption during the pandemic.

To assess whether this is just a situational occurrence or a long-lasting preference, we asked the respondents who had experienced online medical care about their expectation to continue telemedicine after the pandemic is over and the Covid-19 exposure risk is no longer a threat. The results indicate that 32% answered "yes" and 44% answered "maybe". On the opposite hand, only 24% answered "no" implying that they were either not satisfied with their received care or did not expect to be offered telemedicine by doctors after the pandemic. Patients' satisfaction with telemedicine is very important for it can increase the incorporation of online medical care and save clinicians' valuable drive-time for home-visiting palliative care. Moreover, telemedicine can also provide specialty cares for those who live in areas without access to such care, both domestically and internationally (Rockwell et al., 2020).

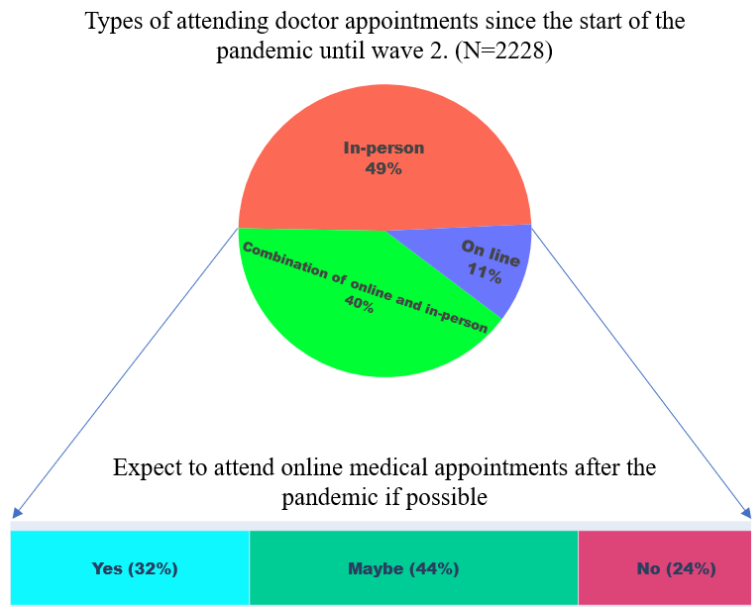

**Figure 6. Distribution of different types of medical appointments during the pandemic and the expectation to use telemedicine after the pandemic**.

### 4.2. Commute Trips and Mode Choice

In the first few months of the COVID-19 pandemic, the world witnessed a drastic change in the transportation sector, with eerie photos of empty streets becoming the hallmark of this era. The pandemic has changed commute patterns through telecommuting, unemployment, and the perceived risk of using transit or other shared services (Harris and Branion-Calles, 2021).

As a part of the survey, we asked pre-covid workers to specify their main travel mode to work and the findings are presented in fig 7 and table 3. Before the pandemic, around 72% of the workers relied on private vehicles to get to their destination. The following most frequently used modes were transit and walk, accounting for transporting around 11% and 3% of the commuters, respectively. In wave 1, the share of private vehicle commuters plummeted to around 40% (i.e., 44% reduction) primarily for shifting to telecommuting and partly for becoming unemployed. As shown in fig. 7, in wave 1, 13% of pre-covid workers lost their jobs while most of them shifted to the not commuter category, making it the biggest class in this period. From wave 1 through wave 2, the percentage of private vehicle commuters increased by 13%, whereas the non-commuter category decreased by 7%.

Regarding the post-covid perceptions, around 66% of the respondents expected to use a private vehicle for commuting, indicating a 9% reduction compared to the pre-pandemic period. Moreover, around 19% of respondents have reported they expected not to commute after the pandemic, which is in line with our findings presented in section 4.1.

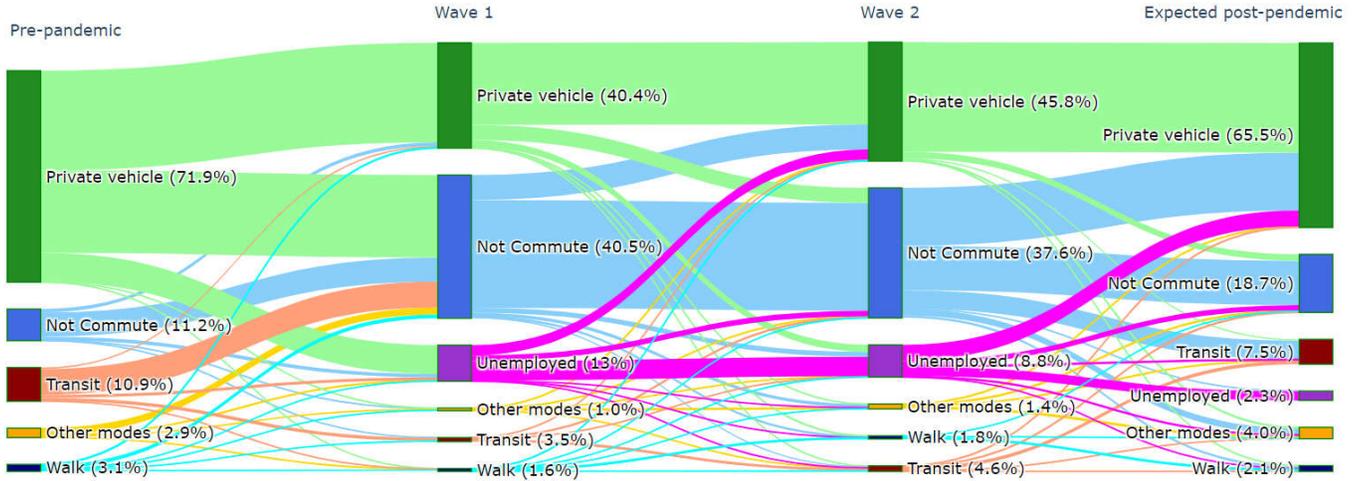

**Figure 7. Proportions and transitions of different categories of workers based on main travel mode to work from pre-pandemic to post-pandemic.**

**Table 3. Percentage change and significance in workers' commute mode from pre-pandemic to post-pandemic.**

| Mode | Change in percentage (p-value) | | | |
|---|---|---|---|---|
| | Pre-pandemic → Wave 1 | Wave 1 → Wave 2 | Wave 2 → Post-pandemic | Pre-pandemic → Post-pandemic |
| Private vehicle | −44% *** (0.000) | +13% *** (0.011) | +43% *** (0.000) | −9% *** (0.000) |
| Transit | −68% *** (0.000) | +31%*** (0.002) | +63%*** (0.000) | −31% *** (0.000) |
| Walk | −48% *** (0.000) | +13% (0.5) | +17% (0.13) | −32%** (0.013) |

Experiencing a 68% reduction, transit commuters' share went through the most massive drop from pre-pandemic to wave 1. Passing the height of the pandemic restrictions, the percentage of transit commute increased by 31% from wave 1 to wave 2. Post-pandemic, our results indicate that although the transit share will continue to grow, it will still be significantly lower than the pre-pandemic period. In other words, the percentage of transit commuters will still be 32% less than before the COVID-19 outbreak.

Furthermore, the percentage of walking commuters sharply declined by 48% from pre-pandemic to wave 1. The fraction of this category then grew by 13% from wave 1 to wave 2; however, this change is not significant. Concerning the post-covid period, our data indicate that 32% fewer commuters are expected to use walking mode to commute to work compared to the pre-pandemic period.

The share of commute modes should be considered in tandem with the frequency of commuting to account for future changes across days of a week. For example, two individuals might report their main commute modes as transit and private vehicle, respectively. However, the former person commutes twice, whereas the latter one commutes five times per week. Since fig. 7 is insufficient to capture this heterogeneity, we asked our respondents to report the number of commute days to work. According to the results depicted in fig. 8, before the pandemic, the average number of commute days was 4.1 days per week, which then plummeted to 1.75 and 1.87 days per week in wave 1 and wave 2, respectively. Regarding the after-pandemic period, it is expected that, on average, people commute 3.42 days per week (i.e., 17% decline compared to the pre-pandemic). Scrutinizing the four periods presented in fig 8, clear distinctions between the distributions of the number of commute days in pre-covid versus post-covid can be observed. Accordingly, the before-pandemic distribution was comprised of mostly workers commuting five days per week with the significantly smaller portion who did not commute. In comparison, the post-covid distribution is different in two ways. First, the share of 5 days per week commuters has decreased while the share of non-commuters has increased. Second, although even after the pandemic most workers expect to commute around five days per week, there is a lump in the post-covid distribution around three days per week, which is unprecedented in the other three distributions. This signifies that in the post-covid period, it is expected that a considerable portion of workers will switch to a hybrid work model with a few days working from home and a few days commuting to an office.

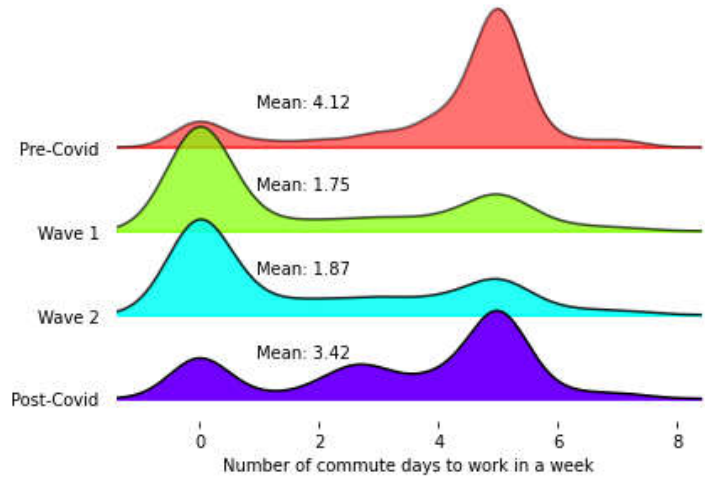

**Figure 8. The number of commute days to work in pre-pandemic, wave 1, wave 2, and post-pandemic periods.**

The increased share of WFH and hybrid workers has significant ramifications on people's travel behavior and subsequently on the transport network (Hensher et al., 2021a). It is worth mentioning that although evidence from the literature suggests that telecommuting can mitigate traffic congestion as well as emission production (Shabanpour et al., 2018), the extent to which the pre-pandemic studies can be generalized to the post-pandemic should be carefully scrutinized. For instance, not going to work reduces the number of commute trips, however, it can provide time flexibility for workers and lead to the generation of other types of trips or even different travel patterns instead of the traditional morning/afternoon peak. As another example, an individual who used to commute 45 min to work from 9 to 5, might assign their previous commute time to go to a gym and benefit from a less crowded environment in the mornings. Besides, he/she potentially can reschedule to finish parts of their job at night and execute other activities (e.g., having lunch with a friend) during the day which by the way translates to the generation of *new* trips.

To have a more comprehensive projection toward the future, we asked all respondents (i.e., not just commuters) to specify their anticipated post-covid use of private vehicle and transit modes (for all purposes and not limited to just commute) by drawing a comparison with the pre-covid period. The reported results are illustrated in fig. 9, indicating that 12% of the people expect to use private vehicles less than before, whereas 17% expect to use them more. This finding supports the idea that although the proportion of private vehicle *commute* trips are expected to decrease, the total use of private vehicles including for *other purposes* might increase. This can be attributed to the flexibility provided by working from home and assigning the saved commute time to other activities. However, this area demands future studies to investigate the different aspects of travel demand in the emerging post-pandemic era.

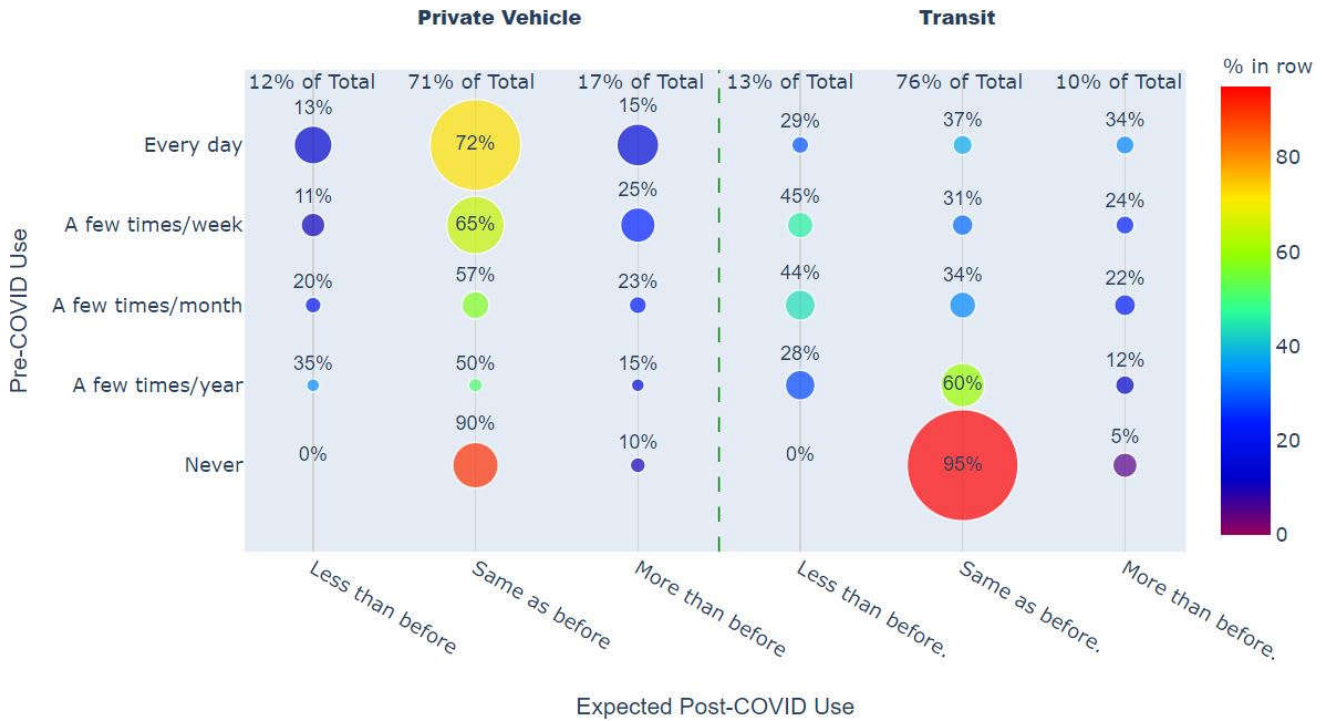

**Figure 9. Pre-covid vs post-covid users of private vehicle and transit mode for all purposes. Please note that the percentage in each cell refers to the share in its corresponding row which is also consistent with the used color scale.**

Moreover, this does not imply that we anticipate either more or less congested streets. Conclusions about congestion require having the data on both volumes of trips as well as the time of day at which trips occur. Hypothetically, a higher number of auto trips can be observed in parallel with less congestion if the trips are more spread throughout the day. This is especially important in the post-covid era when more telecommuters have higher time flexibility to schedule their other non-commute trips for less congested hours. Furthermore, any reduction in congestion due to telecommuting may lead to induced demand, wherein drivers are attracted by the now-less-congested roads.

Regarding the transit mode, 13% of the respondent expect to use it less than before whereas 10% expect to use it more. Shedding light on the underlying components that negatively influence the use of transit mode, we asked the people who want to use this mode less than before, to select among a set of provided factors. The number one concern manifested to be the "I no longer feel safe comfortable sharing space with strangers" by being selected by 73% and 63% of the respondents in wave 1 and wave 2, respectively. This finding is also consistent with the literature that transit mode is perceived as the riskiest travel mode due to the highly contagious nature of the COVID-19 virus (Rahimi et al., 2021). Moreover, WFH was the second most reported reason affecting transit use, which has interestingly increased from 30% in wave 1 to 45% in wave 2.

## .1. Online Shopping

The increased consumers' willingness to shop online to avoid coronavirus in tandem with enforced social distancing regulations has accelerated the expansion of the already growing e-commerce industry to an unprecedented pace. This section strives to understand how people's shopping behaviors have changed by distinguishing between the two categories of grocery and non-grocery items and see whether these changes will persist after the pandemic is over or will bounce back to the pre-pandemic situation. As part of the survey, we asked our respondents to indicate the frequency of using online shopping before, during (i.e., in waves 1 and 2), and after the pandemic (i.e., expected). The results are plotted in fig. 10 and table 4, encompassing three categories of "frequent" (i.e., more than once/week), "infrequent" (i.e., between once/week and once/month), and "rare" (i.e., less than once/month or never) shoppers.

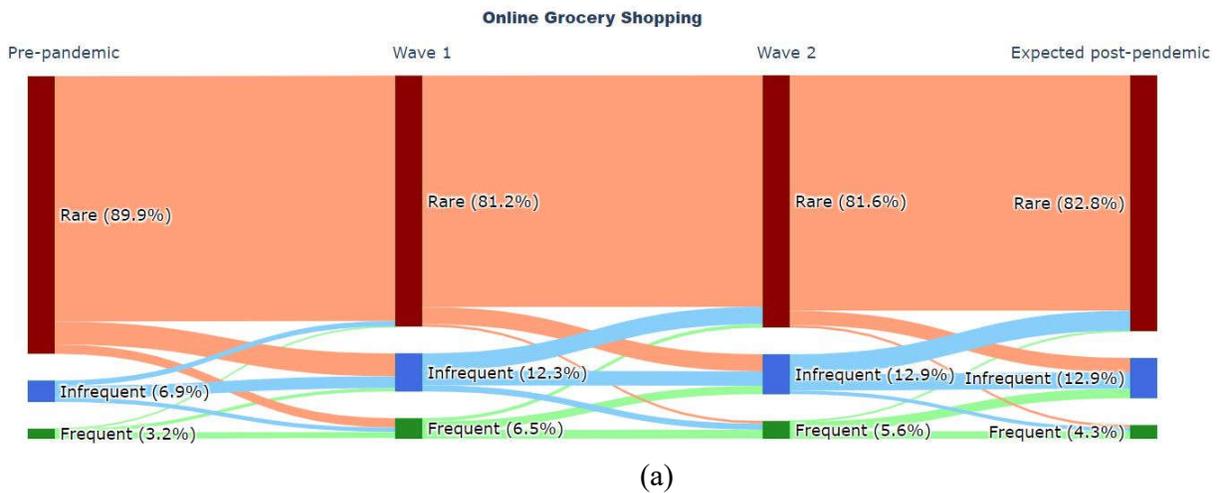

(a)

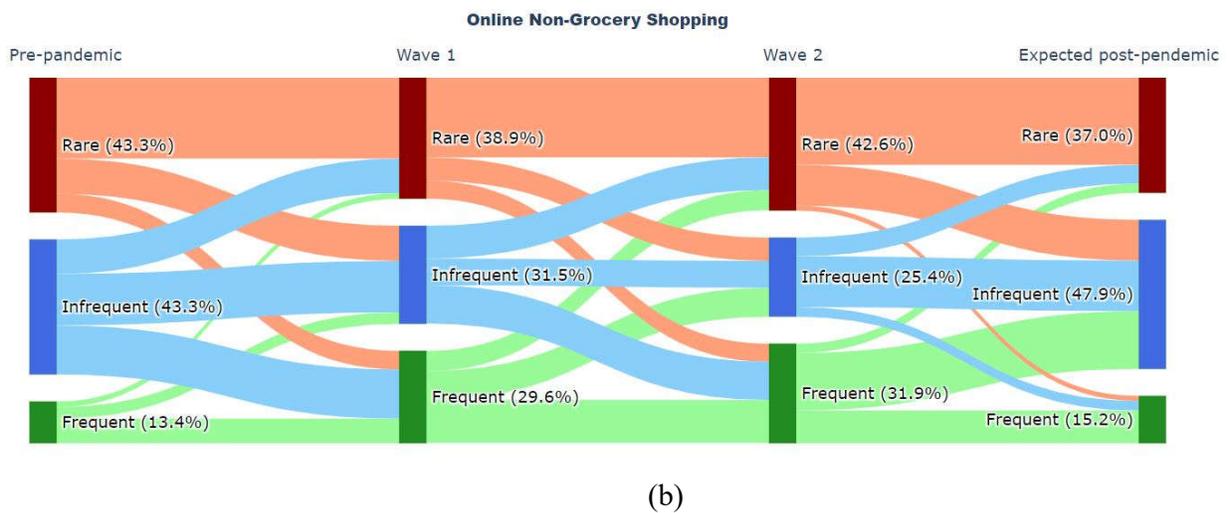

(b)

Figure 10. Proportions and transitions of different categories of online grocery (a) and non-grocery (b) shoppers from pre-pandemic to post-pandemic. The "Frequent" refers to doing online shopping more than once/week, "Infrequent" refers to between once/week and once/month, and "Rare" refers to less than once/month or never.

With respect to grocery items, only around 10% of individuals had utilized online services at least once/month before the pandemic. This proportion then increased to around 19% in wave 1, indicating 90% growth which is also sustained in wave 2.

Furthermore, the frequent online grocery shoppers have almost doubled in wave 1, and although they have decreased in wave 2, this decline is not significant. We also witnessed 78% growth in "infrequent" online grocery shoppers from pre-pandemic to wave 1. Scrutinizing the transitions during the pandemic (i.e., from wave 1 through wave 2), no significant shift can be observed, suggesting the stickiness of the pandemic-induced shopping behavior changes. Although the mentioned statistics manifest a significant surge in online grocery shopping, still most people (i.e., over 80%) have rarely (or never) used online tools for this purpose, neither do they expect to do so after the pandemic. This underscores the importance of taking health safety measures in grocery establishments more seriously as they probably will remain crowded and susceptible to spread the virus.

Table 4. Percentage change and significance in different categories of grocery/non-grocery online shoppers' adoption from pre-pandemic to post-pandemic.

| Commodity type | Online shopping level | Change in percentage (p-value) | | | |
|---|---|---|---|---|---|
| | | Pre-pandemic → Wave 1 | Wave 1 → Wave 2 | Wave 2 → Post-pandemic | Pre-pandemic → Post-pandemic |
| **Grocery** | Frequent | +103% *** (0.000) | −15% (0.058) | −23% *** (0.002) | +34% *** (0.005) |
| | Infrequent | +78% *** (0.000) | −5% (0.48) | 0.0% (1) | +87% *** (0.00) |
| | Rare | −10% *** (0.000) | +0.5% (0.57) | +1.5% (0.51) | −8% *** (0.000) |
| **Non-Grocery** | Frequent | +121 *** (0.000) | +8% ** (0.034) | +52% *** (0.000) | +13% ** (0.014) |
| | Infrequent | −27 *** (0.000) | −19% *** (0.000) | +88% *** (0.000) | +11% *** (0.000) |
| | Rare | −10 *** (0.000) | +10% *** (0.000) | −13% *** (0.000) | −15% *** (0.000) |

*** Indicates a significance level of 99%.
** Indicates a significance level of 95%.

Our respondents were more familiar with online non-grocery purchases than the grocery items, with around 57% reported being frequent or infrequent consumers in the pre-pandemic period. Looking at transitions, frequent shoppers became more than double (i.e., +121% growth) from pre-pandemic through wave 1. Besides, the frequent consumers continued to expand from wave 1 through wave 2 by 8%, further substantiating the consumers' willingness to stick to online shopping even with looser restrictions. However, the infrequent category shrinks simultaneously, primarily because of shifting to the frequent category in waves 1 and 2. The last column in table 4 draws a comparison between the respondents' expectations toward their future shopping choices and their pre-pandemic situation. Accordingly, changes in all three classes of both grocery and non-grocery shoppers are significant, with higher expected growth rates for the grocery market. The biggest expansion is expected to occur in infrequent grocery buyers (i.e., 87%) followed by frequent grocery buyers (i.e., 34%).

These results indicate that consumers' adaptation to online shopping is likely to persist after the Covid-19 passes, which necessitates adopting new strategies and policies to maintain business performance. This conclusion also suggests that the pandemic, that started with sending panicked customers rushing to retail stores, turned out to be the catalyst for digital sales that have transformed the market structure, maybe forever (Kim, 2020).

Recognizing the contributing factors behind consumers' online shopping behavior could potentially assist marketing agencies and urban policymakers to better plan and prepare for the post-pandemic future. Fig. 11 illustrates positive aspects of online shopping that encourage people to shop more in the post-covid situation.

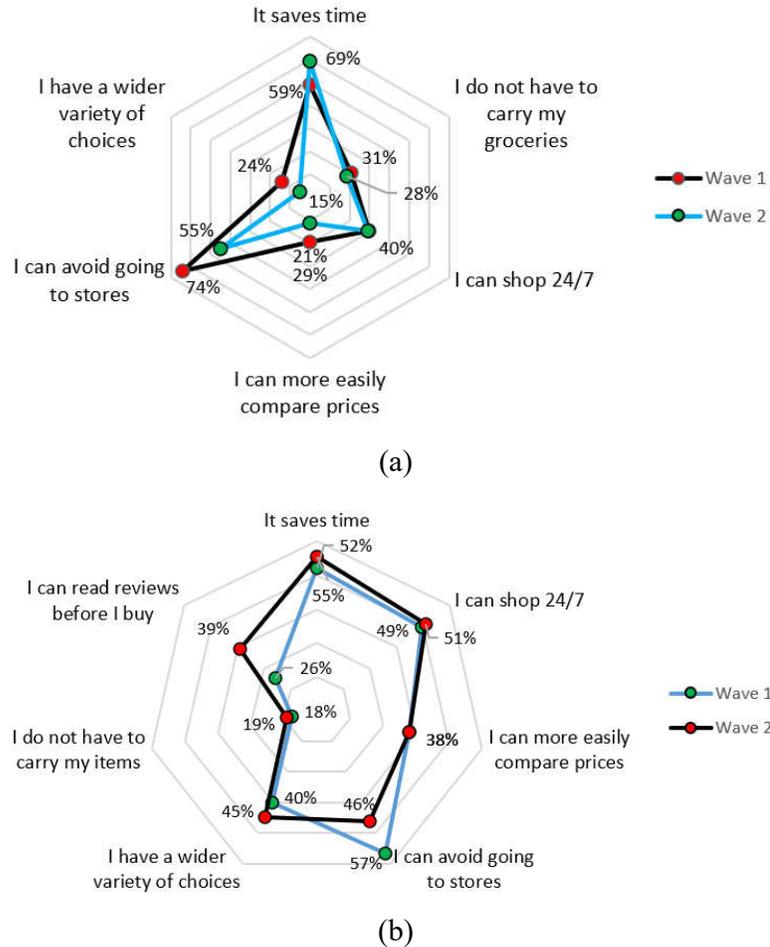

(a)

(b)

**Figure 11. Different factors positively influencing more online grocery (a) and non-grocery (b) online shopping reported in wave 1 and wave 2.**

Our respondents reported "I can avoid going to store" as the most important factor for both online grocery (selected by 74%) and non-grocery (selected by 57%) purchases in wave 1. Following that, "it saves time" and "I can shop 24/7" are the second and third reported reasons favoring online grocery/non-grocery shopping. As the public sensitivity toward the perceived risk of in-store shopping diminished, "it saves time" became the most influential positive factor in wave 2, followed by avoiding going to stores and 24/7 availability.

## .2. Air Travel

According to the WHO Timeline-Covid-19, it only took 10 weeks from the first reported case in Wuhan on 31 Dec 2019, until the outbreak was declared a pandemic on 11 March 2020 (World Health Organization, 2020). This quick spread of Covid-19 across the globe happened mainly due to the virus transmission through air travels which then resulted in massive restrictions that tremendously impacted the demand. According to Transportation Security Administration (TSA), the average daily number of U.S. air travel during June 2019 was 2555626, which dropped to 491835 in June 2020 (i.e., 81% reduction) and to 1890429 in June 2021 (i.e., 26% reduction). To better understand the impacts of the pandemic on long-distance air travel, we designed a set of questions in our survey by distinguishing between leisure/personal and business travel purposes. As illustrated in fig. 12, compared to the pre-pandemic, 24% of respondents expect less while 17% expect more leisure/personal air travels in the post-pandemic period.

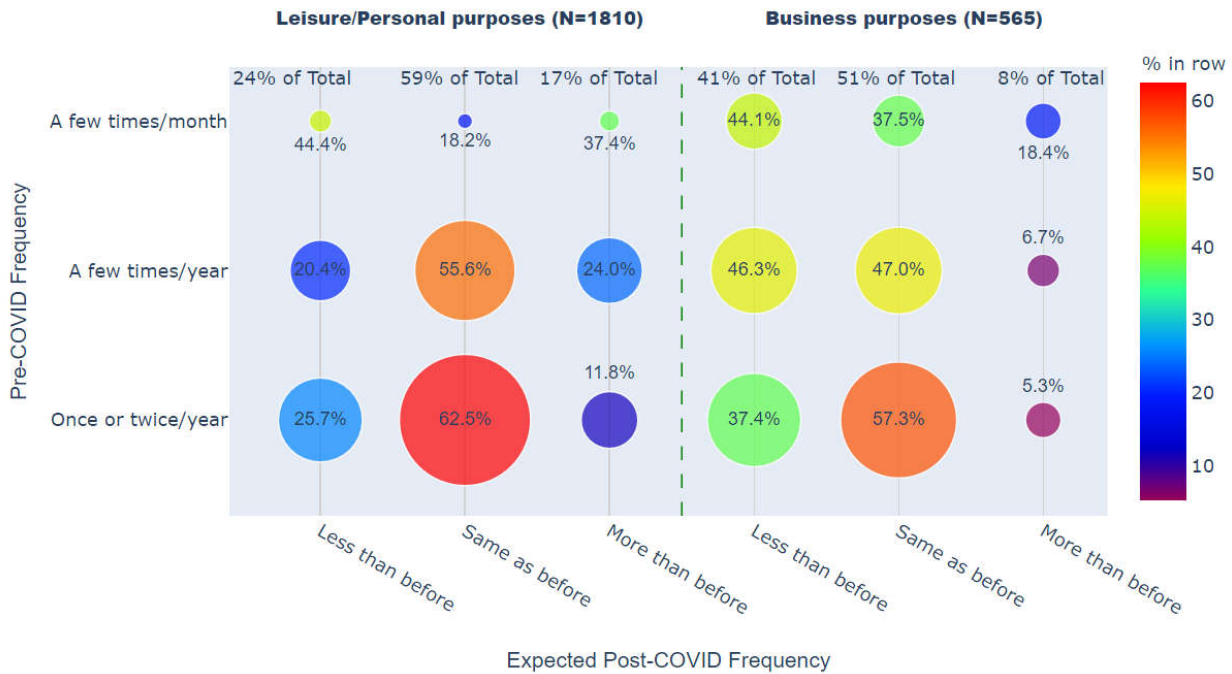

**Figure 12. Pre-covid vs (expected) post-covid frequency of long-distance air travel for "leisure/personal" and "business" purposes. Please note that the percentage in each cell refers to the share in its corresponding row which is also consistent with the used color scale.**

Correspondingly, 41.% of respondents expect less while only 8% anticipate more business air travels in the post-pandemic period. Fig. 12 also clearly indicates the dominant expectation of less business air travels among all three categories of pre-pandemic travelers (i.e., once or twice a week, a few times a week, and a few times a month pre-covid travelers).

The significantly different recovery patterns for leisure/personal and business air travels can be attributed to both the lower resilience of business travels against disruptions (Borko et al., 2020) and also the realization that videoconferencing can substitute a lot of work meetings. To investigate underlying factors affecting peoples' preference to take less air travel, respondents were asked to select among a set of potential contributing reasons (fig. 13). Regarding leisure/personal travel, the first negative factor turned out to be "being uncomfortable sharing a close space with a stranger" (selected by 69% and 53% in waves 1 and 2, respectively).

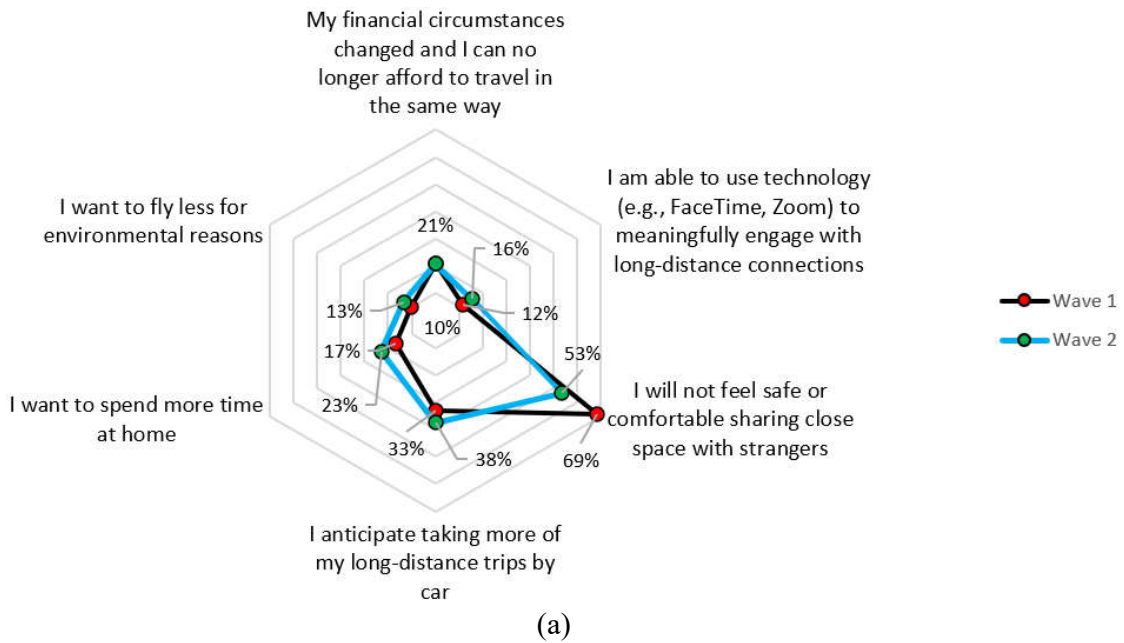

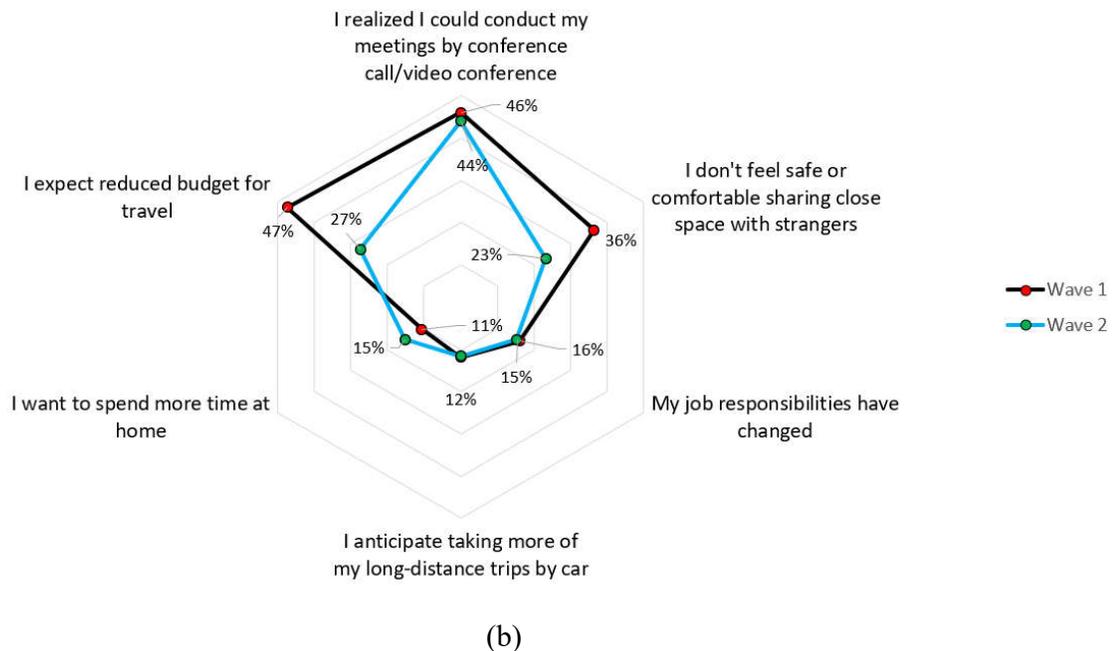

Figure 13. Different factors negatively influencing leisure/personal (a) and business (b) air travels in wave 1 and wave 2.

On the other hand, the most influential factors in business air travel are reported as "I expect reduced budget for travel" (selected by 47%) and "I realized I could conduct my meetings by conference call/video conference" (selected by 46%) in wave 1. However, in wave 2, the budget concern becomes less important (selected by 27%), whereas the ability to conduct meetings through videoconference became the number one factor (selected by 44%) encouraging people to take less business air travel. Although the concern about sharing a close space with strangers in leisure/personal travels is likely to fade after the pandemic, the impact of videoconferencing on business travel seems likely to remain present. This can make the recovery of the air travel industry fragile especially knowing that even a small reduction of business air travel could have serious impacts on airlines (Suau-Sanchez et al., 2020).

Regarding the environmental effects, air travel consumes a huge amount of energy which means releasing a colossal amount of carbon dioxide into the atmosphere. Accordingly, reduced air travel can play a significant role in mitigating greenhouse gas emissions which can positively affect global climate.

- **Conclusions and Policy Implications**

This study presented descriptive/inferential analyses to investigate activity-travel behaviors and attitudes before, during, and after the Covid-19 pandemic. Relying on concrete evidence from nationwide-representative panel data, our results confirm substantial observed and expected changes during and after the pandemic in telecommuting, commute mode choice, online shopping, and air travel habits and preferences. Most likely, the resulted impacts will stick and transcend the pandemic era by shaping different dynamics and patterns in the workplace, shopping market, and transportation sector. In this section, we briefly summarize the extent to which the pandemic-induced changes can be generalized to the post-covid future as well as putting the results into perspective for planners and policymakers.

One of the most conspicuous changes in our data is the substantial expansion of frequent (i.e., more than once/week) telecommuters. The results suggest that around half of employees expect to have the option to work from home post-pandemic, of which 71% anticipate being frequent telecommuters. Over the course of the pandemic, people started to adjust and more telecommuters reported their productivity was the same as or higher than pre-pandemic ( 60% in wave one vs. 71% in wave two). From one side, WFH can lead to decreased peak hour congestion, lower emissions, and saved office costs (Guyot and Sawhill, 2020). Additionally, WFH can decrease the gender gap in terms of monthly income since telecommuter mothers can increase their work hours while they are close to their families (Arntz et al., 2019). On the other side, WFH brings up management challenges for employers to ensure the productivity of their employees since some management styles simply ares not feasible. In our data, the most influential negative factor in teleworkers' productivity is "More distractions at home". Thus, telecommuting productivity is highly related to the home environment (Shamshiripour et al., 2020). Investing in home offices or even choosing residences that are more conducive to WFHcan improve workers' productivity in the long run. Nonetheless, WFH will remain undesirable for some employees who prefer to go to their office. Even people with homes conducive to WFH might prefer to go to the office a few days a week, and benefit from the social interactions in the traditional office culture (Bloom, 2021). Accordingly, a flexible and hybrid telecommuting approach, rather than a compulsory policy, can optimize the benefits while mitigating the disadvantages of WFH across different individuals (Organisation for Economic Co-operation and Development (OECD), 2020).

Moreover, telemedicine can assist health workers in assessing the severity and progression of diseases while protecting other patients and clinicians from infections. It is a cost-effective approach that can broaden access to specialty services in different regions (Kichloo et al., 2020). According to our survey, only 24% of people who experienced telemedicine during the pandemic are not planning to repeat it, whereas the rest hold either a positive (32%) or a neutral (44%) expectation toward online medical appointments after the pandemic. A well-grounded approach including proper triage of patients and distinguishing between those who are suitable for receiving online care versus those who are more proper to be visited in person, educating people about the effectiveness and safety of telemedicine, and improving the telecommunication technology tools can increase the public trust to telemedicine (Portnoy et al., 2020).

The remarkable shift to WFH will also significantly alter the commute travel patterns. Despite the gradual rebound of transit ridership, it is still expected that after the pandemic, transit commuters will be 32% less, compared to the pre-pandemic situation. Moreover, private vehicle commuters are also expected to decrease by 9%. In addition to the reduction in auto/transit commuters, the number of commute days to work is expected to decrease by 17% (i.e., from 4.1 days a week in pre-pandemic to 3.4 days a week in post-pandemic) indicating that a considerable fraction of respondents are going to adopt a hybrid work system. Another notable observation is that 17% of people expect to use private vehicles (for all purposes and not just for commuting) more than before whereas 12% expect to use them less (i.e., compared to the pre-pandemic period). Altogether, these findings indicate that the future of urban mobility includes a lower number of commute trips, higher car dependency, significant transit ridership lost, new and additional generated trips, and different traffic patterns rather than just the conventional morning/afternoon peak. To mitigate the decline in transit ridership, implementing additional health measures such as providing a sufficient amount of sanitizers (Labonté-Lemoyne et al., 2020), screening the temperature of boarding patrons, and avoiding crowdedness can help to rebuild the trust in the transit system. Promoting active and micro-mobility modes is another strategy to move toward a sustainable, resilient, and environment-friendly transportation system with low virus exposure risk (Hosseinzadeh et al., 2021). As one measure, more road space or exclusive lanes can be allocated to active modes to encourage and facilitate their adoption.

Furthermore, preparing for traffic management and transportation planning in the emerging post-pandemic world necessitates a more in-depth focus on occurring changes in people's travel diaries. More reliance on private vehicles and increased time flexibility of telecommuters can create different peak hours throughout the day or even exacerbate congestions in unexpected hours. Further data collection and modeling are crucial since pre-pandemic models and simulations cannot simply predict emerging traffic patterns. Collecting data on people's daily travel patterns provides valuable insights that assist transportation planners in many ways, like manipulating the demand by spreading the demand throughout the day or even between days of a week. For instance, encouraging alternate workdays (e.g., traveling to the office only three days a week) can be consistent with the hybrid work model. Accordingly, all workers are not compelled to commute every day and the commute traffic can be spread over different days (Thombre and Agarwal, 2021).

Regarding the online shopping behavior, our findings reveal that both categories of grocery and non-grocery items are likely to sustain most of their increased share of the market; however, the rise in grocery items is more considerable. The most significant growth is expected to be in the infrequent (i.e., between once/month and once/week) grocery shoppers (87%) followed by frequent (i.e., more than once/week) grocery shoppers (34%) after the pandemic is over. According to our survey, a wide range of policies can be implemented to improve the quality of the online shopping experience for customers. For example, enhancing the reliability of purchases so the customers can be sure that they get exactly what they have ordered, designing user-friendly online platforms for all population groups, making sure that customer reviews are accurate and not based on paid reviews, and providing fast and affordable services to deliver fresh grocery items can help to sustain the growth of the online market. Regardless of the remarkable surge in online grocery shopping, more than 80% of the respondents remained "rare" shoppers (i.e., do online shopping less than once a month) during the pandemic and do not plan to use online shopping more in the future. This statistic highlights the vital role of retail grocery stores in providing people's daily needs even after the pandemic. Thus, authorities should take implementing health safety measures such as sanitizing, ventilation, and air cleaning in grocery establishments seriously to lower the exposure risks. They also need to have contingency plans in case the Covid-19 variants aggravate the situation and social distancing becomes more imperative again. Some studies even have gone further by suggesting booking a time for onsite grocery shopping to ensure social distancing (Bohman et al., 2021).

Lastly, our data indicate that long-distance air travel is still far from recovery. 41% of pre-covid business travelers expect fewer flights (after the pandemic) while only 8% anticipate more, compared to pre-pandemic. Although the demand for leisure/personal air travel is also expected to decline, the decline seems marginal, and recovery seems likely to happen quicker. Airlines can take several actions to mitigate pandemic-induced financial difficulties. With lower business-class trips demand, airlines can focus on reconfiguring airplane cabins by assigning more space to leisure trips. Moreover, airlines can reallocate their fleet to cargo and freight transportation to compensate for their lost revenue (Bouwer et al., 2021).

Another critical point is that moving toward a resilient airline industry is of vital importance after the experience of the Covid-19 outbreak and its catastrophic repercussions. For example, designing airplanes and airports in such a manner that makes implementing social distancing easier, can boost the resiliency of the airline industry.

**Author Contribution:**

Conceptualization: MJ, ER, MM, AD, AM, DS, MWC
Data collection: All authors
Methodology: MJ, ER, MM, MD, AM
Visualization: MJ
Analysis and interpretation of results: MJ, ER, MM, AD, AM, DS, MWC
Writing – original draft: MJ, TM
Writing – review & editing: All authors

**Acknowledgments**

This research was supported in part by the National Science Foundation (NSF) RAPID program under grants no. 2030156 and 2029962, awarded to the University of Illinois at Chicago and Arizona State University. Also, this study was supported by the Center for Teaching Old Models New Tricks (TOMNET), a University Transportation Center sponsored by the U.S. Department of Transportation through grant no. 69A3551747116, as well as from the Knowledge Exchange for Resilience at Arizona State University. This COVID-19 Working Group effort was also supported by the NSF-funded Social Science Extreme Events Research (SSEER) network and the CONVERGE facility at the Natural Hazards Center at the University of Colorado Boulder (NSF Award #1841338) and the NSF CAREER award under grant no. 155173. Any opinions, findings, and conclusions or recommendations expressed in this material are those of the authors and do not necessarily reflect the views of the funders.

**Declaration of competing interest**

None.